\documentclass[cameraready]{Interspeech}

\title{Activation Steering for Accent Adaptation in Large Audio Language Models}

\author[affiliation={1},equalcontribution]{Jinuo}{Sun}
\author[affiliation={1},equalcontribution]{Yang}{Xiao}
\author[affiliation={1}]{Sung Kyun}{Chung}
\author[affiliation={1}]{Qiuchi}{Hu}
\author[affiliation={2}]{Gongping}{Huang}
\author[affiliation={1}]{Eun-Jung}{Holden}
\author[affiliation={1}]{Ting}{Dang}

\address{$^1$
    School of Computer Science, The University of Melbourne, Australia \\
    $^2$
    Wuhan University, China
}

\email{jinuosun@gmail.com, yxiao9550@student.unimelb.edu.au, ting.dang@unimelb.edu.au}


\usepackage{comment}

\usepackage{caption} 
\usepackage{subcaption} 
\usepackage{cite}
\usepackage{soul, xcolor, graphicx}

\definecolor{myblue}{RGB}{200,220,235}

\begin{document}

\maketitle

\begin{abstract}
Accent variability remains a major source of errors in automatic speech recognition, yet most adaptation methods rely on parameter fine-tuning without understanding where accent information is encoded. We treat accent variation as an interpretable subspace in hidden representations and investigate whether it can be identified and controlled directly in activation space. We extract layer-wise encoder activations and estimate mean-shift directions capturing accent-induced representation shifts. By injecting these directions into individual layers and measuring how they align accented and standard embeddings, we derive a layer-wise accent sensitivity profile, revealing that accent information concentrates in a narrow band of middle encoder layers. Leveraging this structure, we further introduce parameter-free accent steering that modifies representations during inference without updating model weights. 
Experiments across eight accents show consistent word error rate reductions. 
\end{abstract}

\section{Introduction}
Accent variability remains a persistent challenge for automatic speech recognition (ASR) systems~\cite{ASR_in_diverse_accents}. Systematic differences in phoneme realization, prosody, and phonotactic patterns across regional and non-native accents often lead to recognition errors that disproportionately affect certain speaker populations~\cite{vctk, l2}. In real-world deployments, including voice assistants, call centers, and educational technologies, such performance disparities degrade user experience and raise important concerns regarding fairness and accessibility~\cite{koenecke2020racial}. Despite advances in large-scale pretraining and robust acoustic modeling~\cite{xiao2025dg}, recognition gaps across accents continue to be observed.

Conventional accent adaptation techniques typically rely on supervised fine-tuning, accent-specific modeling, or data augmentation~\cite{ghorbani2022domain,qian2022layer,prabhu2023accented,shi2026affectcodec}. While effective in traditional ASR settings, these approaches become increasingly costly and operationally restrictive in the era of foundation-scale speech models. Full-parameter adaptation is computationally expensive and may compromise generalization across diverse accents and tasks~\cite{poor_generalization, unseen_gap_1, unseen_gap_2}. As modern speech systems are frequently deployed as shared, large-scale backbones, there is growing need for lightweight and scalable adaptation mechanisms.

Recent large audio language models (LALMs), such as Whisper~\cite{radford2023robust}, combine powerful acoustic modeling with large language modeling to achieve strong zero-shot generalization. While Parameter-Efficient Fine-Tuning (PEFT) methods and bottleneck adapters~\cite{houlsby2019adapters,hu2021lora,learn_and_dont_forget, parameter-efficient_model_reprogramming} offer a more lightweight adaptation approach by adding only a small number of trainable parameters on top of a frozen base model, these approaches still optimize the added parameters heuristically~\cite{song2024lora_whisper, prasad2024lowrank_asr,mas-lora, xu2024towards}. Without explicitly localizing or constraining updates to layers and subspaces that are most sensitive to accent-related variation, these methods may perform unnecessary adaptation in accent-insensitive regions of the model and risk entangling accent compensation with higher-level semantic representations, limiting both efficiency and control.

Understanding how accent variation is organized in representation space is critical for designing scalable and controllable adaptation strategies. If accent corresponds to identifiable subspaces in hidden activations, then targeted representation-level interventions may offer an alternative to parameter-intensive fine-tuning~\cite{xiao2026adapting}.
Recent studies in large language models show that high-level attributes, such as sentiment, style, or safety~\cite{Category-wise_safety_steering, Affine_concept_editing, dynamic_steering,turner2024activation,zou2023representation}, can align with approximately linear directions in activation space. These directions, often referred to as steering vectors, are typically constructed as mean activation shifts between contrasting concepts~\cite{rimsky2024steering,ilharcoediting}. When added to hidden states at selected layers, they nudge representations toward desired regions of the space, enabling controllable behavioral modulation at inference time without modifying model parameters. This perspective motivates the central question of this work: \emph{Does accent variation correspond to a structured and controllable subspace in LALMs?}



In this paper, we first conduct a layer-wise analysis of hidden activations to understand the geometric structure of accent variations. This investigation serves two primary purposes. From an \emph{interpretability} perspective, it reveals how the model distributes accent information across layers. Furthermore, from a \emph{controllability} perspective, it determines whether these accent features are organized enough to allow direct adjustments. 
Based on this foundation, we construct steering vectors to adjust accent representations during model inference. Our method inserts learned accent directions into specific hidden states. We subsequently evaluate their effectiveness in automatic speech recognition (ASR) tasks across diverse pronunciations. 
To the best of our knowledge, this is the first study to systematically analyze and apply vector steering for accent-robust automatic speech recognition. Our extensive experiments on the VCTK and L2-ARCTIC datasets demonstrate the middle-to-late layers are the most sensitive to accent-induced representation shifts. Moreover, mean shift steering significantly improves the word error rate across eight distinct accents without updating model parameters. We observe that interventions in the middle layers provide the most effective control. 
This study offers a principled and scalable way to reduce accent-induced recognition disparities in speech foundation models, advancing both robustness and fairness in real-world ASR deployment.

\section{Accent Subspace Analysis and Steering}

\subsection{Layer-wise Accent Subspace Analysis}
We analyze how accent-related variation is organized across encoder layers and quantify how strongly each layer responds to accent-specific representation shifts. This yields a layer-wise sensitivity profile that highlights which layers are most suitable for subsequent accent steering.

\subsubsection{Accent Representation Perturbation}
We construct text-matched utterance pairs to isolate accent-related acoustic variation from linguistic content. For each target accent, we create two types of pairs: (i) \textit{cross-standard-accent} pairs $(x_s, x_a)$, where $x_a$ is an accented utterance and $x_s$ is a standard-English utterance with an identical transcript, and (ii) \textit{within-single-accent} pairs $(x_a, x_a)$ formed by different speakers from the same accent group. The cross-standard-accent pairs capture systematic accent-induced differences, while the within-single-accent pairs serve as a control to account for speaker-specific factors such as timbre or prosody, helping us separate accent effects from general inter-speaker variation. Given accented speech samples ${x_a}$ and standard speech samples ${x_s}$, our layer-wise analysis then quantifies, at each encoder layer, how the latent representations of accented speech differ from those of standard English, and how effectively a layer supports aligning these two clusters. Concretely, for each layer $l$, we first extract the token-level hidden activations $\mathbf{H}^{(l)}$. For each speech sample $i$, we apply mean pooling over time for comparing across layers to obtain an utterance-level representation $\bar{\mathbf{h}}_{i}^{(l)}$. 
Based on these pooled representations, we compute an accent mean-shift direction from a source accent group $s$ to a target accent group $a$:
\vspace{-2mm}
\begin{equation}
\mathbf{d}_{s \rightarrow a}^{(l)}
=
\frac{1}{|G_s|}\sum_{j \in G_s}\bar{\mathbf{h}}_{j}^{(l)}
-
\frac{1}{|G_a|}\sum_{i \in G_a}\bar{\mathbf{h}}_{i}^{(l)} 
\label{eq:mean_shift}
\vspace{-2mm}
\end{equation}
where $G_s$ and $G_a$ denote the sample sets corresponding to the standard and accent speech, respectively. This vector $\mathbf{d}_{s \rightarrow a}^{(l)}$ characterizes the accent shift between the two representation clusters at layer $l$. To probe the sensitivity of layer $l$ to this accent direction, we apply a controlled perturbation to its hidden activations:
\vspace{-2mm}
\begin{equation}
\tilde{\mathbf{H}}^{(l)} = \mathbf{H}^{(l)} + \alpha \cdot \mathbf{d}_{s \rightarrow a}^{(l)},
\vspace{-2mm}
\end{equation}
where $\alpha$ is optimized to $1.0$, corresponding to a one-unit accent mean-shift perturbation, and $\mathbf{d}_{s \rightarrow a}^{(l)}$ is broadcast across all time steps of $\mathbf{H}^{(l)}$. The resulting perturbed activations $\tilde{\mathbf{H}}^{(l)}$ are then propagated through the remaining encoder and projector layers. 

\subsubsection{Quantifying Layer Sensitivity via Accent Alignment Score (AAS)}
\label{sec:within_accent_control}
Given the layer-wise perturbation defined above, 
we quantify how much each layer contributes to reducing the representation gap between accented and standard speech.
Our analysis is conducted on the audio encoder of Qwen2-Audio-7B~\cite{qwen2audio}, which consists of 32 Whisper-style encoder layers followed by a multi-modal projector. Since accent variation is primarily expressed through acoustic patterns, we focus on hidden-state transformations within the audio encoder and measure their propagated effects in the projector space, which provides the final speech representation passed to the downstream language model.

For each source-target speech pair, we apply the mean-shift displacement $\mathbf{d}_{s \rightarrow a}^{(l)}$ at audio encoder layer $l$ and measure its propagated effect in the multi-modal projector space. 
Because the projector output remains time-dependent, we further apply mean pooling over the projector time steps to obtain a single utterance-level representation for each sample, on which cosine similarity is computed.

We define the Accent Alignment Score (AAS) as the change in cosine similarity between the perturbed source representation and the target representation:
\vspace{-2mm}
\begin{equation}
            \mathrm{AAS}_{s \rightarrow a}^{(l)} = \cos\!(\tilde{\mathbf{z}}_{\mathrm{proj}}^{\,a, (l)}, \mathbf{z}_{\mathrm{proj}}^{\,b}) - \cos\!(\mathbf{z}_{\mathrm{proj}}^{\,a}, \mathbf{z}_{\mathrm{proj}}^{\,b})
            \vspace{-3mm}
\end{equation}
Here, $\tilde{\mathbf{z}}_{\mathrm{proj}}^{\,a,(l)}$ denotes the projector output of the source speech sample after applying the layer-$l$ perturbation. Meanwhile, $\mathbf{z}_{\mathrm{proj}}^{\,a}$ and $\mathbf{z}_{\mathrm{proj}}^{\,b}$ represent the baseline projector outputs for the source and target speech samples, respectively. A positive $\mathrm{AAS}$ indicates that the layer-wise perturbation moves the source representation closer to the target accent representation.

We apply the same procedure to both cross-standard-accent and within-single-accent pairs, and compute their AAS values at each layer. Specifically for within-single-accent pairs, we calculated a within accent mean-shift vector $\mathbf{d}_{a \rightarrow a}^{(l)}$ between two speakers of the same accent, following the same AAS calculation procedure.

To isolate the additional alignment gain observed in cross pairs beyond general speaker variation, we compute the specificity score, which measures the amount of the patching effect that truly comes from the accent:
\vspace{-2mm}
\begin{equation}
\mathrm{Spec}^{(l)}
=
\begin{aligned}
\overline{\mathrm{AAS}}_{\mathrm{cross}}^{(l)}
-
\overline{\mathrm{AAS}}_{\mathrm{within}}^{(l)}
\end{aligned}
\vspace{-3mm}
\end{equation}
where $\mathrm{Spec}^{(l)}$ denotes \textit{Specificity} score at layer $l$, and $\overline{\mathrm{AAS}}$ denotes the average score across all evaluated speech pairs in their respective sets. The subscript $\mathrm{cross}$ denotes cross-standard--accent pairs, while $\mathrm{within}$ denotes within-accent pairs. A positive $\mathrm{Spec}^{(l)}$ value indicates that the patching effect at this specific layer is indeed caused by accent differences. Consequently, it is not caused by other speaker-level variations. Based on this logic, the layer sensitivity score \(\mathrm{Sensitivity}^{(l)}\) is defined as \(\max(0, \mathrm{Spec}^{(l)})\)
which is used for ranking across layers. In addition, our layer-wise analysis adopts a bidirectional design (\textit{standard} $\rightarrow$ \textit{target} and \textit{target} $\rightarrow$ \textit{standard}) to reduce direction-specific bias and provide a more robust estimate of layer sensitivity. Each direction is computed independently and then averaged.

\subsection{Inference-Time Accent Steering}
The layer-wise sensitivity analysis identifies candidate layers for controllable accent intervention. Building on the analysis, we next construct steering vectors to modulate accent-related representation during inference. We then test whether steering at these layers can produce measurable shifts that improve the downstream ASR performance of accented speech toward that of standard speech. To extract the generalized steering vector and avoid the effect of speakers and sentences, we design an \textit{extraction set} and an \textit{evaluation set}. They are isolated from the speakers and texts. This design ensures that the steering vector is estimated from data that shares neither speaker identity nor text content with the evaluation data.

For steering, we reuse the mean-shift direction defined in \eqref{eq:mean_shift}, but compute it only from the extraction set. Before injection, we normalize the direction $\mathbf{d}_{s \rightarrow a}^{(l)}$ to unit norm:
\vspace{-2mm}
\begin{equation}
\hat{\mathbf{d}}_{s \rightarrow a}^{(l)}
=
\frac{\mathbf{d}_{s \rightarrow a}^{(l)}}{\left\lVert \mathbf{d}_{s \rightarrow a}^{(l)} \right\rVert}
\vspace{-2mm}
\end{equation}
This separates direction from magnitude and makes the steering strength parameter $\alpha$ comparable across layers. During inference, we inject the normalized steering vector into the hidden states at the selected layer $l$:
\vspace{-2mm}
\begin{equation}
\tilde{\mathbf{H}}^{(l)}
=
\mathbf{H}^{(l)} + \alpha \hat{\mathbf{d}}_{s \rightarrow a}^{(l)}
\vspace{-3mm}
\end{equation}
where $\alpha$ controls the steering strength. The vector $\hat{\mathbf{d}}_{s \rightarrow a}^{(l)}$ is broadcast across all time steps of $\mathbf{H}^{(l)}$. We implement this intervention using a forward hook, so no model parameters are modified. We propose that the vector of an accent direction estimated from a subset of speech data can generalize effectively to unseen speakers and unseen utterances. The evaluation is in the Results section.

\section{Experiment Settings}

\subsection{Datasets and Accents}

\noindent \textbf{Native accents:} We adopt the VCTK dataset~\cite{vctk}, a high-fidelity corpus featuring speakers with diverse regional accents, to study native English variations. Specifically, we select Scottish, South African, Canadian, Irish, and Northern Irish accents for our experiments.
As VCTK includes a standard English group, we utilize this subset directly as our reference for comparison.

\noindent \textbf{Non-native accents:} We utilize the L2-ARCTIC corpus~\cite{l2}, which provides manual phonetic-level annotations of speech from diverse linguistic backgrounds. In this study, we focus on Hindi, Arabic, and Spanish accents.
Since L2-ARCTIC lacks an internal native reference, we draw native speakers from its source, the CMU-ARCTIC dataset~\cite{cmu}, which consists of phonetically balanced sentences designed for speech synthesis. Because both datasets utilize identical reading scripts, the CMU-ARCTIC data serve as an ideal, matched reference group.


\vspace{-5pt}
\subsection{Pair Construction and Data Splits}

\noindent \textbf{Accent Subspace Analysis:} For each of the above accents, we construct 1000 cross-standard-accent pairs and 500 within-single-accent pairs. No data split is required.

\noindent \textbf{Inference-Time Accent Steering:} For each of the above target accents, we adopt a strict data splitting protocol: $80\%$ of the speakers are assigned to the extraction set, while the remaining $20\%$ are reserved for the evaluation set. 
We further enforce no transcript overlap between extraction pairs and evaluation utterances to avoid text leakage. For each accent group, we randomly sample 1,000 speech pairs from the extraction set to estimate the mean-shift vector.

\vspace{-5pt}
\subsection{Evaluation Protocol}

We evaluate steering in a single-layer sweep across 32 encoder layers, where each layer and each steering strength $\alpha$ are tested independently. We choose $\alpha$ = [0.5, 1, 2, 5] to cover different scale of steering strength.
Steering effectiveness is evaluated by the resulting change in Word Error Rate (WER).

To reduce bias from difficulty, the evaluation set is constructed by balanced sampling: We first input the candidate utterances to the trained model. Then we select half of the target utterances from samples with $\mathrm{WER}=0$, and the other half from samples with $\mathrm{WER}>0$. In the default setting, we sample 100 utterances from each group, yielding 200 evaluation utterances in total. We use the same balanced subset and evaluation protocol for the steering results in Sec. 4 and for the Base, Steer, and PEFT comparison in Table 1.

\section{Results and analysis}

\begin{figure*}[t!]
  \centering
  \begin{subfigure}[t]{0.33\textwidth}
    \centering
    \includegraphics[width=\linewidth]{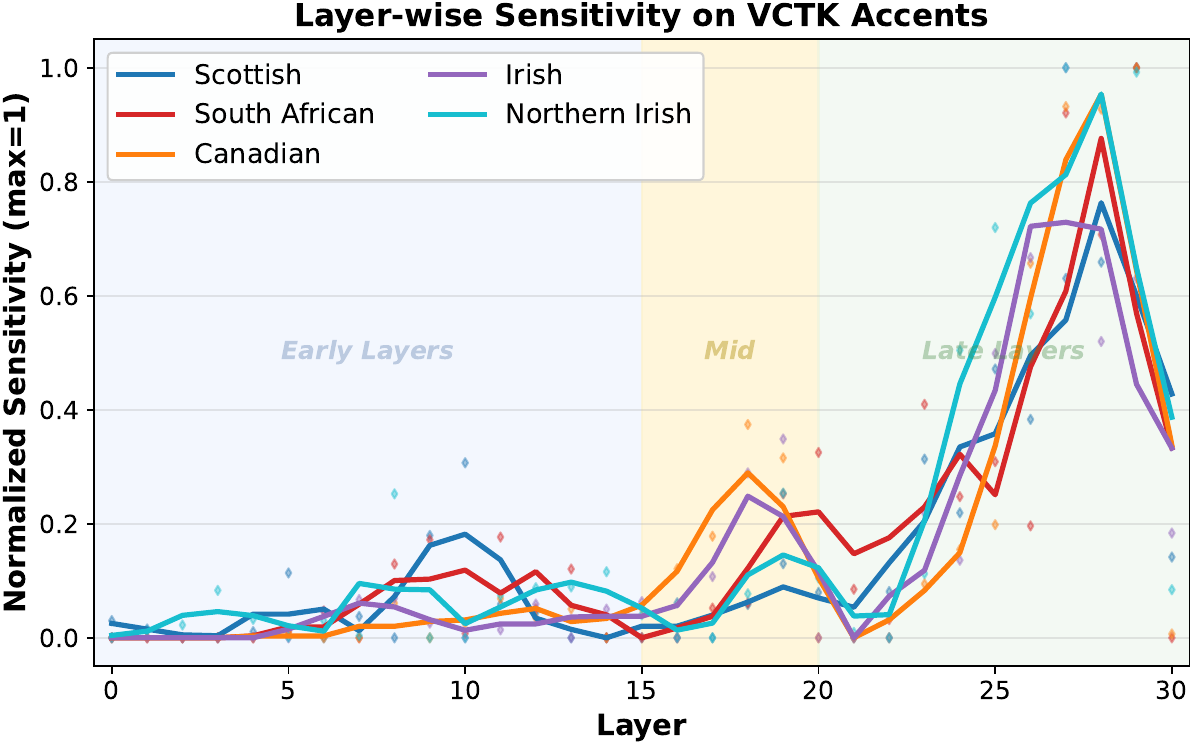}
    \caption{Layer-wise sensitivity (VCTK)}
    \label{fig:sens_vctk}
  \end{subfigure}\hfill
  \begin{subfigure}[t]{0.33\textwidth}
    \centering
    \includegraphics[width=\linewidth]{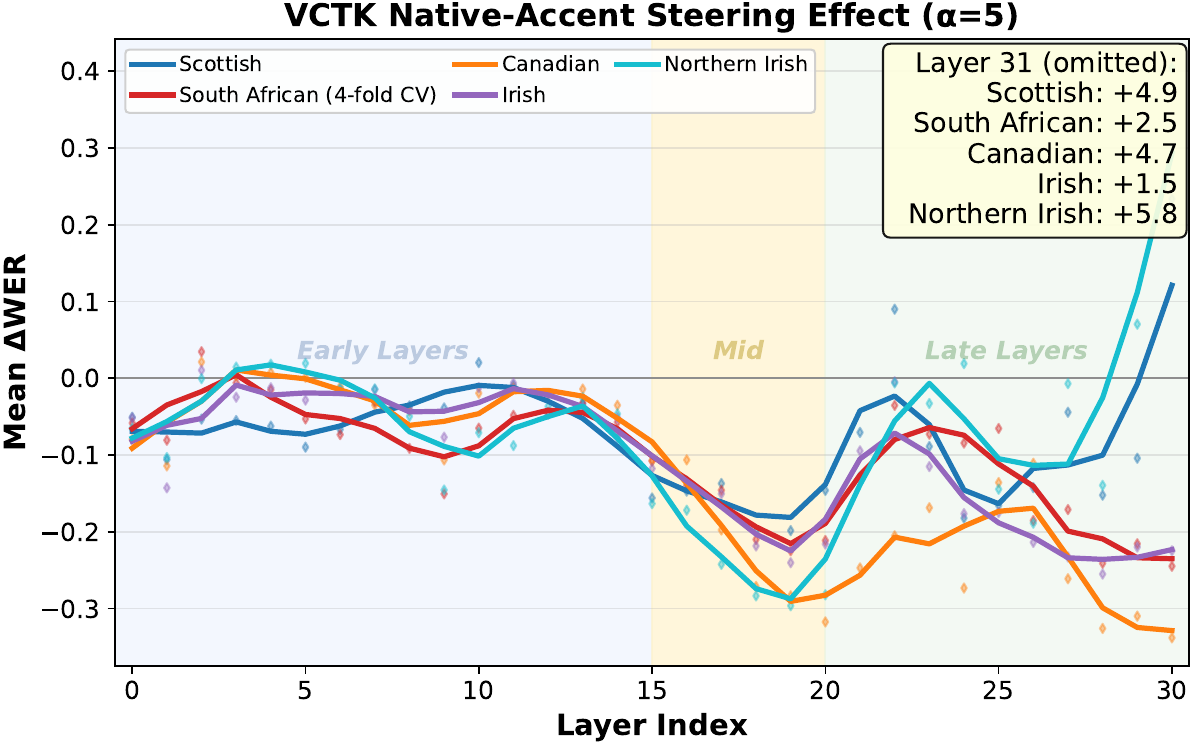}
    \caption{Steering $\Delta$WER (VCTK)}
    \label{fig:steering_vctk}
  \end{subfigure}\hfill
  \begin{subfigure}[t]{0.33\textwidth}
    \centering
    \includegraphics[width=\linewidth]{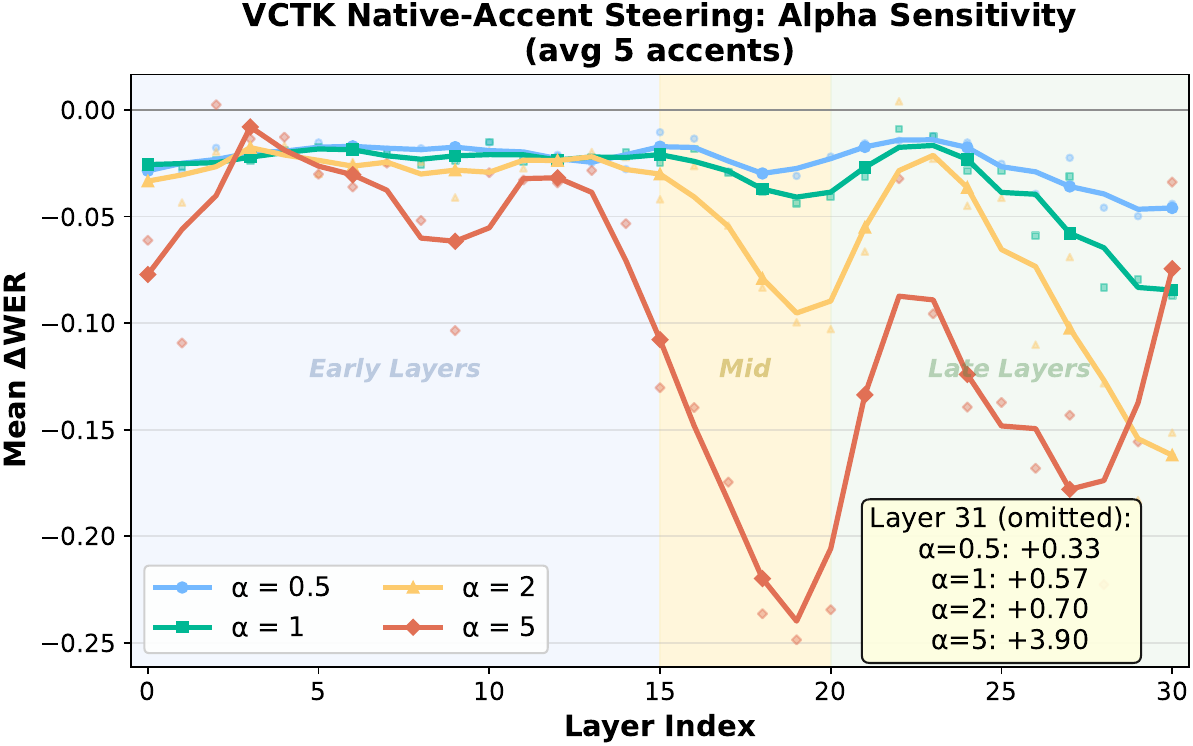}
    \caption{Avg.\ $\Delta$WER vs.\ $\alpha$ (VCTK)}
    \label{fig:alpha_vctk}
  \end{subfigure}

  \vspace{4pt}

  \begin{subfigure}[t]{0.32\textwidth}
    \centering
    \includegraphics[width=\linewidth]{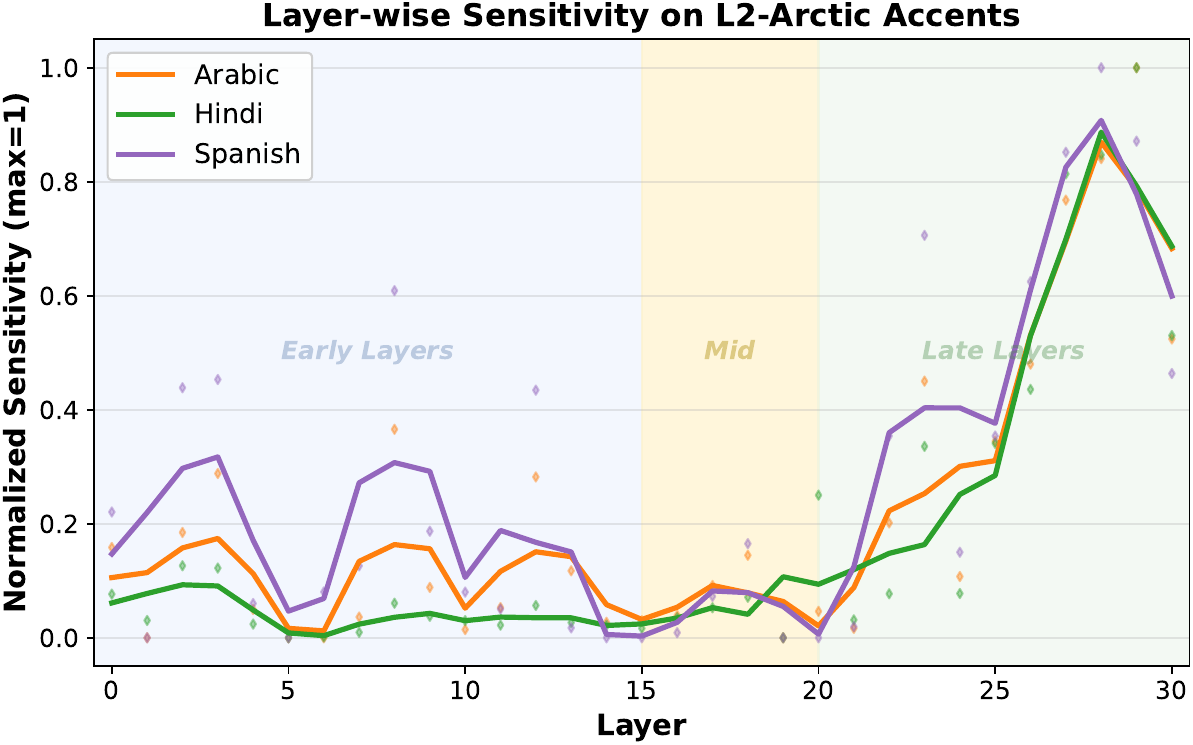}
    \caption{Layer-wise sensitivity (L2-ARCTIC)}
    \label{fig:sens_l2}
  \end{subfigure}\hfill
  \begin{subfigure}[t]{0.32\textwidth}
    \centering
    \includegraphics[width=\linewidth]{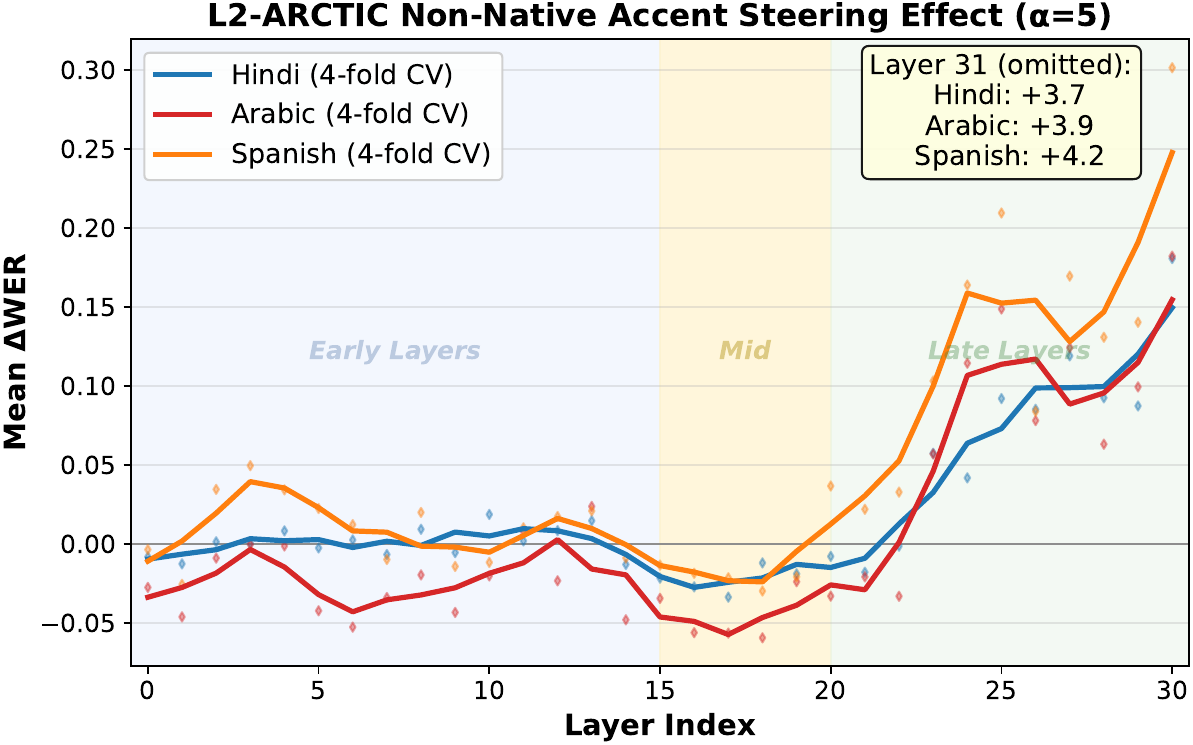}
    \caption{Steering $\Delta$WER (L2-ARCTIC)}
    \label{fig:steering_l2arctic}
  \end{subfigure}\hfill
  \begin{subfigure}[t]{0.32\textwidth}
    \centering
    \includegraphics[width=\linewidth]{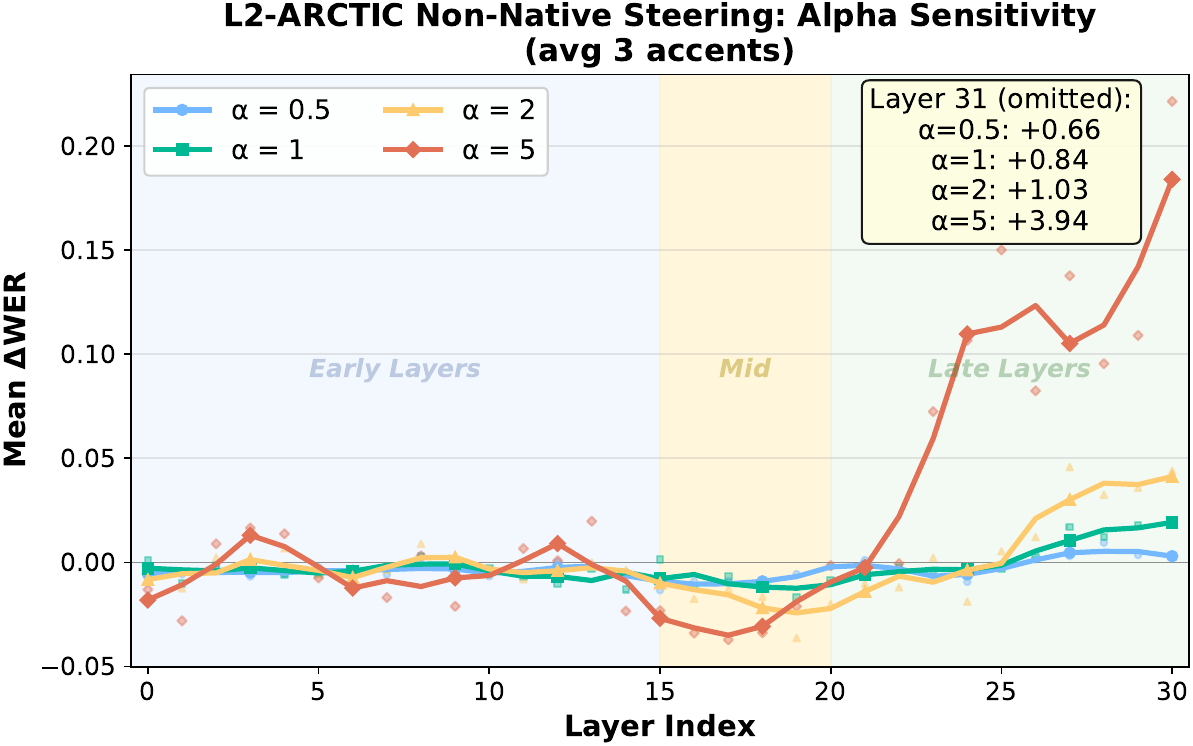}
    \caption{Avg.\ $\Delta$WER vs.\ $\alpha$ (L2-ARCTIC)}
    \label{fig:alpha_l2arctic}
  \end{subfigure}
\vspace{-3mm}
  \caption{Layer-wise accent subspace analysis and steering results with an alpha sweep.
           Top row: VCTK; bottom row: L2-ARCTIC.}
  \label{fig:accent_subspace_steering}
  \vspace{-5mm}
\end{figure*}

\subsection{Accent Subspace Analysis}

To reduce scale differences, each accent group is normalized independently. Layer 31 is excluded as it directly precedes the linear multi-modal projector. Sensitivity analysis on VCTK shows a shared pattern: low sensitivity in early layers (0–14), an emerging peak near layer 15, and a sharp rise to global maximums from layer 21 onward. L2-ARCTIC exhibits similar trends but with stronger global fluctuations, suggesting non-native accents induce more distributed representational changes.
The steering window is divided into early (0–14), middle (15–19), and late (20–30) stages. Early layers process low-abstraction acoustic information~\cite{pasad2021layer,pasad2023comparativ}; native accents show weak sensitivity here, while non-native accents trigger higher peaks due to stronger low-level acoustic deviations. This makes early intervention less controllable. 
Based on this observation, the middle to late layers are identified as more promising target locations for steering vectors.

\subsection{Analysis of Layer-Wise Steering Dynamics}
\subsubsection{Steering Effect on Native Accents}
We first analyze the layer-wise steering effect on the VCTK dataset to understand native accent representations. As shown in Figure~\ref{fig:steering_vctk}, we observe a consistent trend across the Scottish, South African, Canadian, Irish, and Northern Irish groups. In the early layers, the intervention yields minimal impact on the mean word error rate. 
However, the middle layers have the most evident
WER reductions. The strongest gains are concentrated in middle layers, reaching up to about 30\% $\Delta$WER for some accents on the balanced subset. By contrast, late-layer steering is unstable and often harmful.
For example, layer 31 experiences massive error increases across all groups. This geometric pattern suggests that native accent characteristics are primarily encoded in intermediate representations. Therefore, targeted interventions in these middle sections provide optimal control without disrupting higher-level semantic understanding.


\subsubsection{Steering Effect on Non-native Accents}

We extend our analysis to non-native speech using the L2-ARCTIC corpus. We evaluate the mean shift steering effect on Hindi, Arabic, and Spanish variations. Similar to the native setup, the performance changes reveal a distinct spatial structure as shown in Figure~\ref{fig:steering_l2arctic}. The early layers show relatively stable performance with minimal metric fluctuations. Subsequently, the middle layers again demonstrate a clear improvement in recognition accuracy. Although the error reduction is smaller than that of native speech, reaching around 5\% $\Delta$WER drop, the localized trend remains highly consistent. Furthermore, steering in the final layers causes severe degradation. Specifically, layer 31 alone adds approximately 4.0 to the error rate across all three linguistic backgrounds. This structure confirms that non-native accent variations also occupy a controllable subspace. Consequently, applying targeted steering to specific middle layers offers a scalable pathway for adaptive systems.

\subsubsection{Cross-Accent Steering Pattern and Insights}

Empirical results reveal consistent layer-wise patterns for both native and non-native accents. Layers 0–14 are largely unresponsive to steering, yielding minimal metric changes. Conversely, the middle layers (15–19) emerge as the optimal window, showing consistent, significant error reductions across all linguistic backgrounds. This suggests middle-layer representations are not yet fixed, allowing accent perturbations to propagate stably and confirming these intermediate states effectively isolate accent characteristics for controllable steering. Late-layer steering produces highly unstable, divergent results. For non-native accents, it frequently causes representation collapse; even for native accents, gains do not exceed those of the middle layers. This poor controllability likely stems from increasingly fixed representations, leaving too few layers to reorganize injected directions. Finally, terminal layer 31 consistently causes massive performance degradation, confirming the tail layer is unsuitable for injection.

\subsubsection{Steering Strength Analysis}
To compare how different $\alpha$ values affect performance, we compute, for both native and non-native accents, the mean $\Delta$WER steering effect under each $\alpha$ injection strength. Figures~\ref{fig:alpha_vctk} and~\ref{fig:alpha_l2arctic} illustrate the resulting $\alpha$-parameter sensitivity trends. We observe that as $\alpha$ increases, the layer-wise steering effect exhibits larger fluctuations and higher improvement peaks. Notably, once the steering strength exceeds a critical threshold, both native and non-native accents show signs of an accelerated collapse starting from around layer 27.

Specifically, for native accents, $\alpha=2$ yields strong steering effects in the late layers with a generally depth-progressive trend, but performs poorly in the mid layers. In contrast, with $\alpha=5$, the mid-layer performance improves substantially, surpassing the late layers and reaching higher peaks, while the late layers display stronger oscillations and clearer collapse behavior. For non-native accents, the overall trend is more consistent: the mid layers constitute the globally optimal intervention window, whereas the collapse in the late layers becomes increasingly pronounced as $\alpha$ grows, with the collapse onset shifting toward earlier layers. Overall, the $\alpha$ sweep further corroborates that the mid layers (15-19) form a shared and most suitable steering window across both native and non-native accents. We stop the sweep at $\alpha=5$ because this regime already shows clear instability in deeper layers.

\begin{table}[t]
\centering
\scriptsize

\caption{
Comparison of steering and PEFT on the balanced evaluation subsets defined in Sec. 3.3. Lower WER is better. Base denotes the unadapted WER on the balanced subset; Steer and PEFT denote WER after inference-time steering and parameter-efficient fine-tuning on the same subset. $\Delta$WER is reported relative to Base. The Train column reports the number of accent-specific training speech pairs available to PEFT.
}
\vspace{-2mm}
\label{tab:steering_finetune_compare}
\resizebox{\linewidth}{!}{
\begin{tabular}{lcccccc}
\toprule
\toprule
Accent & Train & Base & Steer & PEFT & St.\,$\Delta$ & PEFT\,$\Delta$  \\
\midrule
Scottish   & 197 & 26.72\% &  6.80\% &  9.25\% & \textbf{-19.92\%} & -17.47\%  \\
S. Afr.    &  44 & 29.86\% &  4.35\% & 27.10\% & \textbf{-25.51\%} &  -2.76\%  \\
Canadian   &  51 & 37.27\% &  3.47\% & 32.60\% & \textbf{-33.80\%} &  -4.67\%  \\
N. Irish   &  49 & 36.27\% &  6.64\% & 31.57\% & \textbf{-29.63\%} &  -4.70\%  \\
Irish      &  87 & 31.91\% &  6.41\% & 30.28\% & \textbf{-25.50\%} &  -1.63\%  \\
\midrule
Arabic     & 802 & 18.13\% & 10.07\% &  7.20\% &  \textbf{-8.06\%} & -10.93\%  \\
Hindi      & 790 & 14.26\% & 10.22\% &  7.82\% &  \textbf{-4.04\%} &  -6.44\%  \\
Spanish    & 796 & 15.31\% &  9.39\% &  8.61\% &  \textbf{-5.92\%} &  -6.70\% \\
\bottomrule
\bottomrule
\end{tabular}}
\vspace{-6mm}
\end{table}

\vspace{-5pt}
\subsection{Comparable study with PEFT Baseline}





Table 1 compares the fine-tuning approach with our proposed steering method. Analysis reveals a critical link between training sample size and adaptation efficacy. Fine-tuning excels with large datasets, such as the Arabic, Hindi, and Spanish sets containing approximately 800 training pairs. Conversely, parameter-updating techniques struggle in data-scarce scenarios. For accents with fewer than 100 samples, including South African, Canadian, Northern Irish, and Irish, fine-tuning performance is remarkably poor. In contrast, steering excels under these constraints, achieving 4.04\% to 33.80\% WER reductions (28.3\%–90.7\% relative) across eight accents using very few samples. Furthermore, as steering requires no parameter updates, it preserves original model capabilities, providing an efficient, flexible solution for inference adaptation.

\section{Conclusion}

In this paper, we propose a lightweight method to mitigate accent variability in automatic speech recognition through inference-time activation steering. Initially, our analysis identified the middle layers(15-19) in the audio encoder of Large Audio Language Models as the most stable region for targeted intervention. Based on this geometric insight, we introduced a parameter-free adaptation technique using mean shift steering vectors. Consequently, this approach significantly reduces recognition errors across diverse accents. Furthermore, it vastly outperforms conventional fine-tuning in data-scarce environments without requiring weight updates. Ultimately, this research provides a highly scalable pathway for developing inclusive speech technologies.

\clearpage
\section{Generative AI Use Disclosure}
We use generative AI tools for polishing the manuscript, e.g., correcting the grammar.
\bibliographystyle{IEEEtran}
\bibliography{mybib}

\end{document}